\documentclass[aps,prb,reprint,superscriptaddress,showpacs]{revtex4-1}
\usepackage{graphicx}
\usepackage{bm}

\begin{document}

\title{Optically induced nonreciprocity by a plasmonic pump in semiconductor wires}
\author{Kil-Song Song}
\affiliation{Department of Physics, Kim Il Sung University, Taesong District, 02-381-4410 Pyongyang, Democratic People's Republic of Korea}
\author{Song-Jin Im}
\email{sj.im@ryongnamsan.edu.kp}
\affiliation{Department of Physics, Kim Il Sung University, Taesong District, 02-381-4410 Pyongyang, Democratic People's Republic of Korea}
\author{Ji-Song Pae}
\affiliation{Department of Physics, Kim Il Sung University, Taesong District,  02-381-4410 Pyongyang, Democratic People's Republic of Korea}
\author{Chol-Song Ri}
\affiliation{Department of Physics, Kim Il Sung University, Taesong District,  02-381-4410 Pyongyang, Democratic People's Republic of Korea}
\author{Kum-Song Ho}
\affiliation{Department of Physics, Kim Il Sung University, Taesong District,  02-381-4410 Pyongyang, Democratic People's Republic of Korea}
\author{Chol-Sun Kim}
\affiliation{Department of Physics, Kim Il Sung University, Taesong District,  02-381-4410 Pyongyang, Democratic People's Republic of Korea}
\author{Yong-Ha Han}
\affiliation{Department of Physics, Kim Il Sung University, Taesong District,  02-381-4410 Pyongyang, Democratic People's Republic of Korea}

\begin{abstract}
In most studies on all-optical diodes spatial asymmetry has been by necessity applied to break Lorentz reciprocity. Here we suggest a paradigm for optically induced nonreciprocity in semiconductor wires which are spatially asymmetry-free and provide a very simple and efficient platform for plasmonic devices. An azimuthal magnetic field induced by a plasmonic pump in the semiconductor wire alters the material parameters and thus results in a cross-nonlinear modulation of the plasmonic signal. Peculiarly the nonlinear wavenumber shift has opposite signs for forward and backward signals whereas Kerr or Kerr-like nonlinearity does not break Lorentz reciprocity in spatially symmetric structures. This principle may open an avenue towards highly integrated all-optical nonreciprocal devices. 
\end{abstract}
\maketitle

Surface wave and plasmonics offer the ability to manipulate light on nanoscale and achieve enhanced nonlinear interactions \cite{1Kauranen2012} because it enables local field enhancement and inhomogeneity on nanoscale \cite{2Barnes2003, 3Schuller2010}. Plasmonic waveguides find applications as optical interconnects in highly integrated optoelectronic devices \cite{4Bozhevolnyi2006} and also have been used for significantly enhancing nonlinear optical effects by deep-subwavelength mode confinement combined with a mode propagation length much longer than the wavelength. In particular, plasmonic wires provide a highly efficient platform for plasmonics and metamaterials \cite{5Akimov2007, 6Simovski2012}, and nonlinear optics \cite{7Udabyahaskar2014, 8Marini2011, 9Im2016_2}. The inverse Faraday effect (IFE) has attracted much attention due to its great potential for ultrafast all-optical switching of magnetization \cite{10Kirilyuk2010}. It is widely accepted that magnetoplasmonic structures provide a great potential for enhancing magneto-optical effects \cite{11Armelles2013, 12Temnov2010, 13Belotelov2011, 14Ho2018, 15Pae2018}. This method has been successfully demonstrated to enhance the IFE and downscale the region of the optically-induced magnetization to nanometer scale \cite{16Guyader2015, 17Dutta2017, 18Chekhov2018, 19Ignatyeva2019, 20Kuzmin2016A}. Recently, we reported the IFE induced by surface plasmon polaritons (SPPs) in planar plasmonic waveguides and its reaction to the SPPs manifested by a different type of third-order nonlinearity \cite{21Im2017, 22Im2019, 23Ho2020, 24Ri2019}.

Optical nonreciprocity becomes a topic of strong scientific interest for nanophotonic, quantum-optical, and optoelectronic applications \cite{25Potton2004}. The mechanisms to break Lorentz reciprocity are magneto-optical phenomena \cite{26Wang2008, 27Wang2009, 28Hadad2010, 29Sayrin2015, 30Pae2019}, including topological quantum-Hall effect \cite{26Wang2008, 27Wang2009}. An alternative approach is to break time-reversal symmetry in the system \cite{31Lira2012, 32Sounas2013}. To build all-optical diode, nonlinear effects including opto-mechanical interactions have been exploited \cite{33Fan2012, 34Peng2014, 35Gallo2001, 36Yu2015, 38Fleury2014, Bino2018, Li2020, 37Weiss2010, Shen2016}.  In most studies on optically induced nonreciprocity spatial asymmetry has been by necessity introduced to break Lorentz reciprocity.

In this paper we suggest a paradigm for optically induced nonreciprocity in a spatially asymmetry-free, cavity-free and very simple waveguide structure: semiconductor wire. We note that electrically controlled nonreciprocal propagation in plasmonic wire structures has been suggested \cite{39Davoyan2014, 40Bliokh2018}, while our case is promising for all-optical applications. We also note that it has been suggested that second harmonic generations can be supported also in centrosymmetric particle chains via optically induced magnetic effects alleviating the need for non-symmetric structures \cite{Steinberg2011}. An azimuthal magnetic field induced by a plasmonic pump power flowing in the semiconductor wire alters the material parameters resulting in a cross-nonlinear modulation of the plasmonic signal. The nonlinear wavenumber shift has opposite signs for forward and backward signals whereas Kerr-like nonlinearity does not break Lorentz reciprocity in spatially symmetric structures. This principle may provide a promising route for the development of highly integrated all-optical diode.

We consider a semiconductor wire in a conventional dielectric background as shown in Fig. 1(a). We assumed that the semiconductor permittivity has a negative value Re$(\varepsilon_1)<0$ which is justified at a frequency lower than the plasma frequency of the semiconductor.
The effect of an external magnetic field on surface plasmon in semiconductors has been studied theoretically \cite{41Brion1972} and experimentally \cite{42Palik1976} in the ${1970}^{\rm{th}}$. The electric displacement of the semiconductor under an external magnetic field ${\vec{H}}^{\mathrm{IFE}}$ is expressed as
\begin{eqnarray}
\vec{D}=\varepsilon_0\hat{\varepsilon}\vec{E}=\varepsilon_0(\varepsilon_1\vec{E}+i\alpha\vec{E}\times{\vec{H}}^{\mathrm{IFE}}),
\label{eq1}
\end{eqnarray}
where $\varepsilon_1=\varepsilon_\infty[1-\omega_\mathrm{pl}^2/(\omega\left(\omega+i\gamma_s\right))]$ and 
$\alpha\left(\omega\right)\approx e\mu_0\varepsilon_\infty{\omega_{\rm{pl}}}^2/{(m}_{\mathrm{eff}}\omega^3)$.  $\varepsilon_0$ and $\mu_0$ are the vacuum permittivity and permeability, respectively and $\alpha$ is the magneto-optical susceptibility. $\omega_\mathrm{pl}$ is the plasma angular frequency, $e$ and $m_{\mathrm{eff}}$ are charge and effective mass of electron, respectively, $\varepsilon_\infty$ is the background permittivity, which depends on the properties of the bound electrons in the material. $\gamma_s$ is the effective electron collision frequency, responsible for the material absorption. Here, we assumed that the cyclotron frequency $\omega_c={e \mu_0{H}^{\mathrm{IFE}}}/m_{\mathrm{eff}}$ is far smaller than the considering angular frequency $\omega$.

\begin{figure*}
\centering
\includegraphics[width=0.8\textwidth]{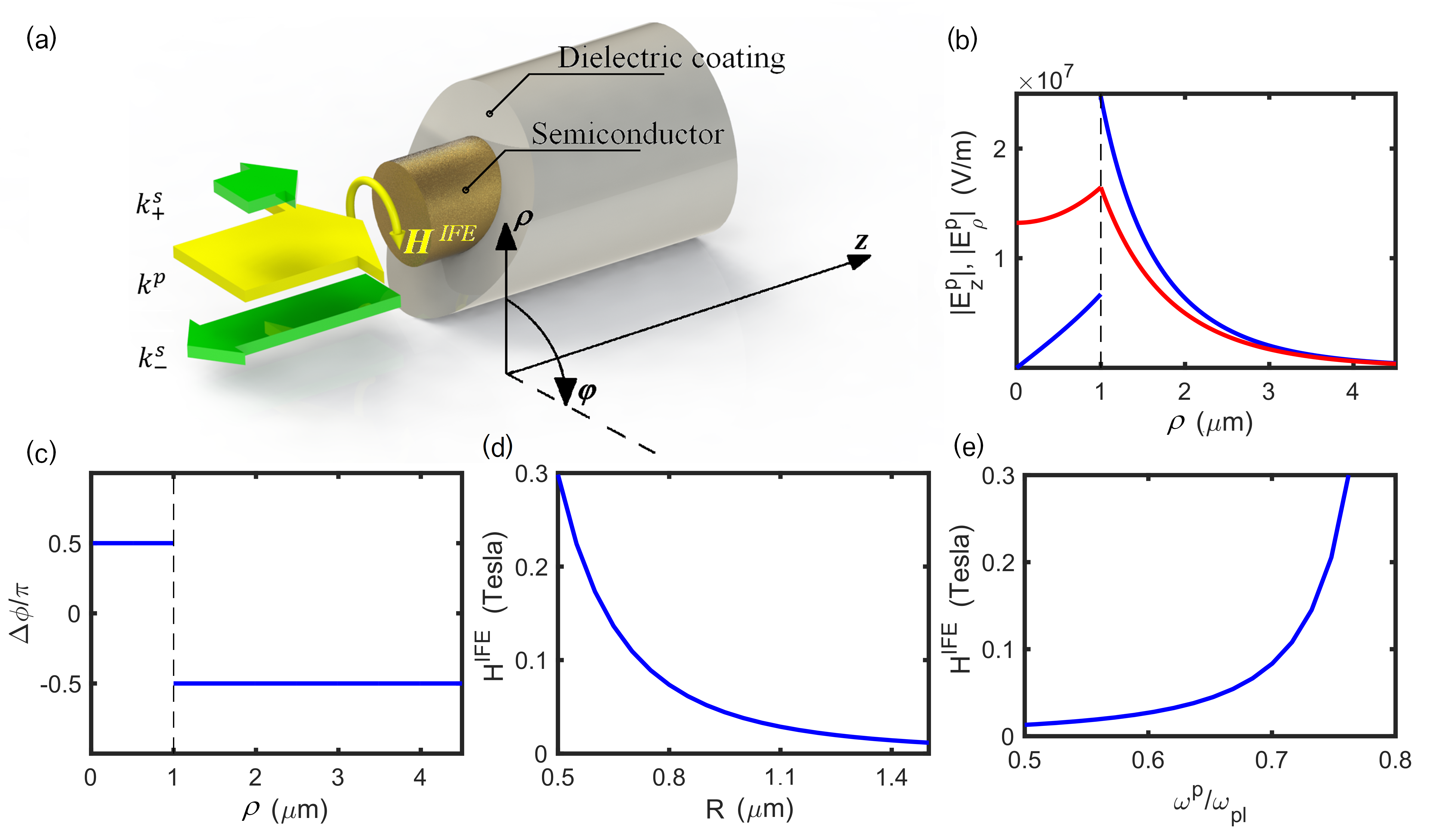}
\caption{(a) Scheme of a nonreciprocal semiconductor wire. (b) Distributions of the $\hat{z}$-axis electric field component $E_z^\mathrm{p}$(the red curve) and the $\hat{\rho}$-axis electric field component $E_\rho^\mathrm{p}$ (the blue curve) of the plasmonic pump in the semiconductor wire. (c) Phase difference between  $E_z^\mathrm{p}$ and $E_\rho^\mathrm{p}$. (d) Strength of optically induced magnetic field versus the wire radius. (e) Strength of optically induced magnetic field versus the pump frequency. Here we assumed a wire radius of 1 $\mu \mathrm{m}$ in (b), (c) and (e), a pump frequency of ${\omega^\mathrm{p}=0.63\omega}_\mathrm{pl}$ in (b)-(d) and a pump power of $P$=1W in (b)-(e). For calculations we assumed parameters of $\varepsilon_\infty=15.68$, $\omega_\mathrm{pl}=3.14\times{10}^{13}s^{-1}$, $m_{\mathrm{eff}}=0.022m_e$ where $m_e$ is electron's mass which correspond to those of the n-type InSb \cite{42Palik1976}.}
\label{fig:1}
\end{figure*}   
\begin{figure*}
\centering
\includegraphics[width=0.8\textwidth]{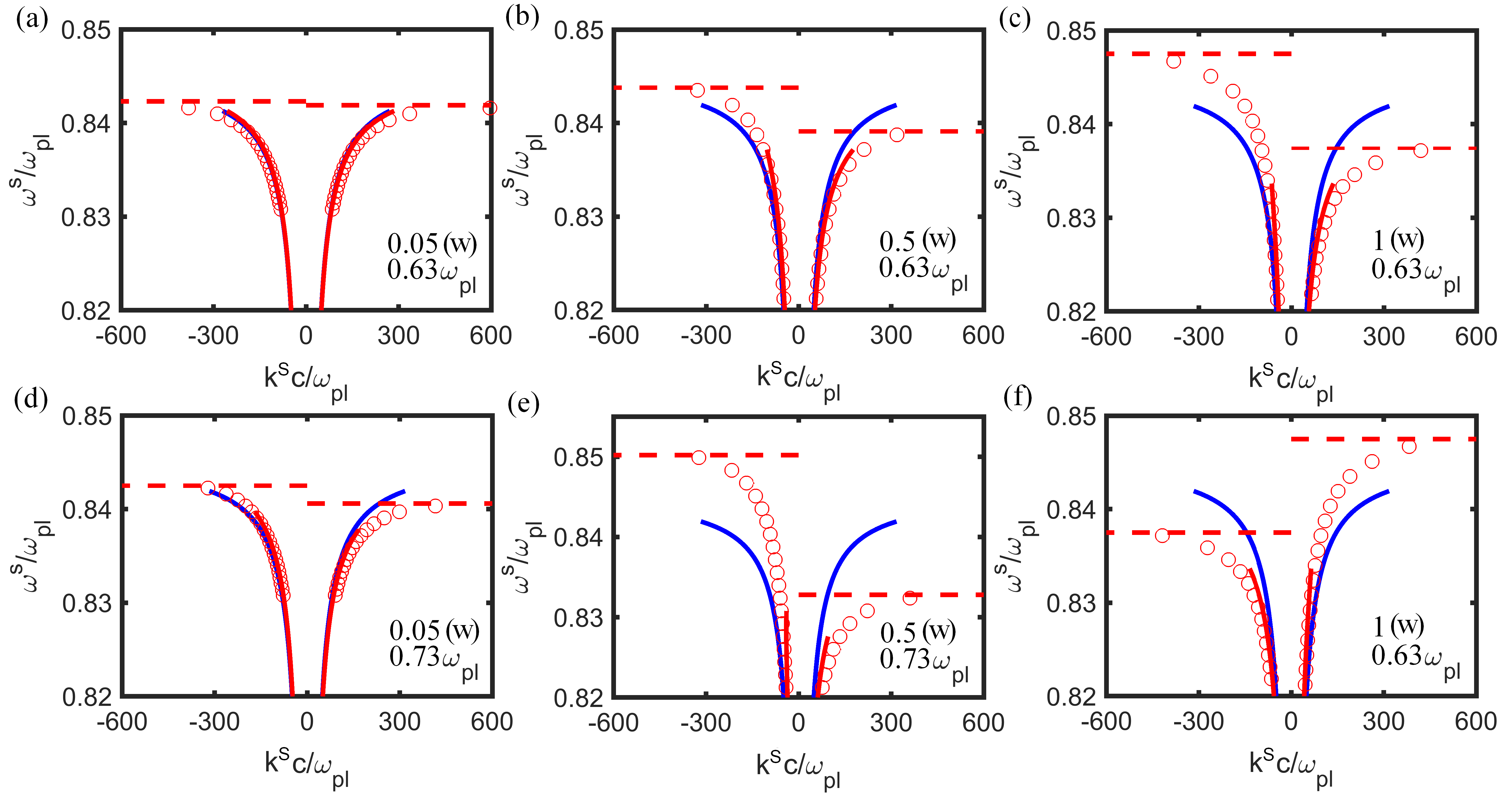}
\caption{All-optically controlled dispersion relation of the plasmonic signal at several characteristic regimes of the mode power and operation frequency of the plasmonic pump. (a) The dispersion relation for the mode power $P$=0.05W and operation frequency $\omega^\mathrm{p}=0.63\omega_\mathrm{pl}$ of the pump. (b) is for $P$=0.5W and $\omega^\mathrm{p}=0.63\omega_\mathrm{pl}$, (c) is for $P$=1W and $\omega^\mathrm{p}=0.63\omega_\mathrm{pl}$, (d) is for $P$=0.05W and $\omega^\mathrm{p}=0.73\omega_\mathrm{pl}$ and (e) is for $P$=0.5W and $\omega^\mathrm{p}=0.73\omega_\mathrm{pl}$. (f) The dispersion relation for $P$=1W and $\omega^\mathrm{p}=0.63\omega_\mathrm{pl}$ for counterpropagating forward pump and backward signal. (a)-(e) are for forward pump and forward signal. Blue and red solid curves denote the dispersion relation in the absence and presence of the plasmonic pump power flow, respectively. Red solid curves have been calculated by using Eq. (6) where the fundamental plasmonic mode has been obtained by numerically solving the dispersion relation in cylindrical coordinates\cite{39Davoyan2014, 40Bliokh2018} in the absence of the plasmonic pump power flow. Red circles are obtained by numerically mode solving the Maxwell equations in the presence of the plasmonic pump power flow. Here we neglected the loss in the wire to display the concept.}
\label{fig:2}
\end{figure*}  

\begin{figure}
\centering
\includegraphics[width=0.3\textwidth]{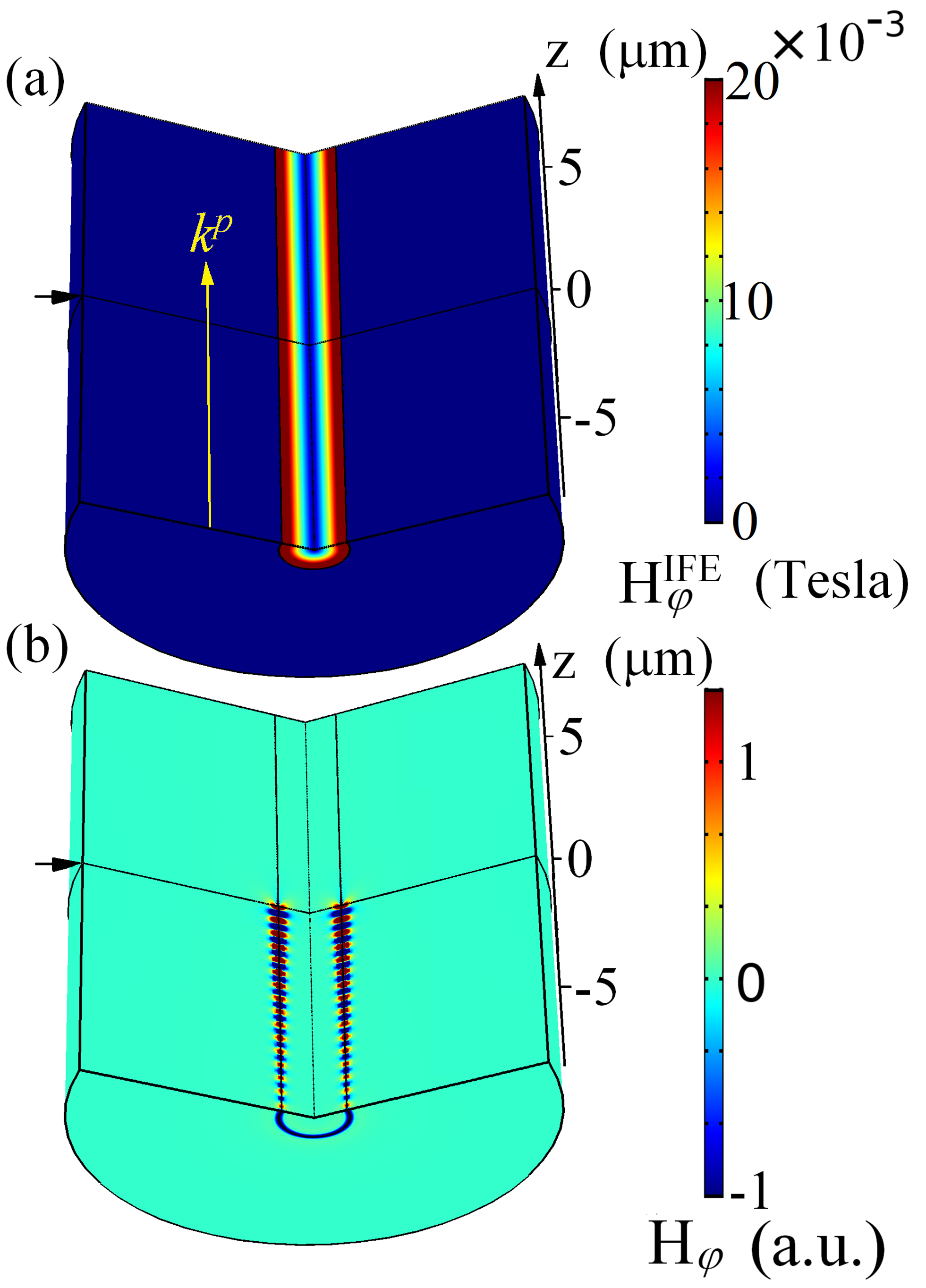}
\caption{Nonreciprocal propagation in the semiconductor wire. (a) Distribution of the static azimuthal magnetic field $H_\varphi^{\mathrm{IFE}}$ induced by the plasmonic pump power flow. (b) Distribution of the azimuthal magnetic field component of the plasmonic signal. The input pump power is 1W, the pump frequency is $\omega^\mathrm{p}=0.63\omega_\mathrm{pl}$ and the signal frequency is $\omega^\mathrm{s}=0.84\omega_\mathrm{pl}$. The corresponding dispersion can be seen in Fig. 2(c). For the simulation, a small loss of $\gamma_s=0.0005\omega_\mathrm{pl}$ in the wire has been assumed. A time-harmonically oscillation of azimuthal magnetic current at z=0 has been introduced to excite both of upward and downward signals. The plasmonic pump has been excited by introducing a power of the fundamental mode at the down port.}
\label{fig:3}
\end{figure}   

\begin{figure}
\centering
\includegraphics[width=0.3\textwidth]{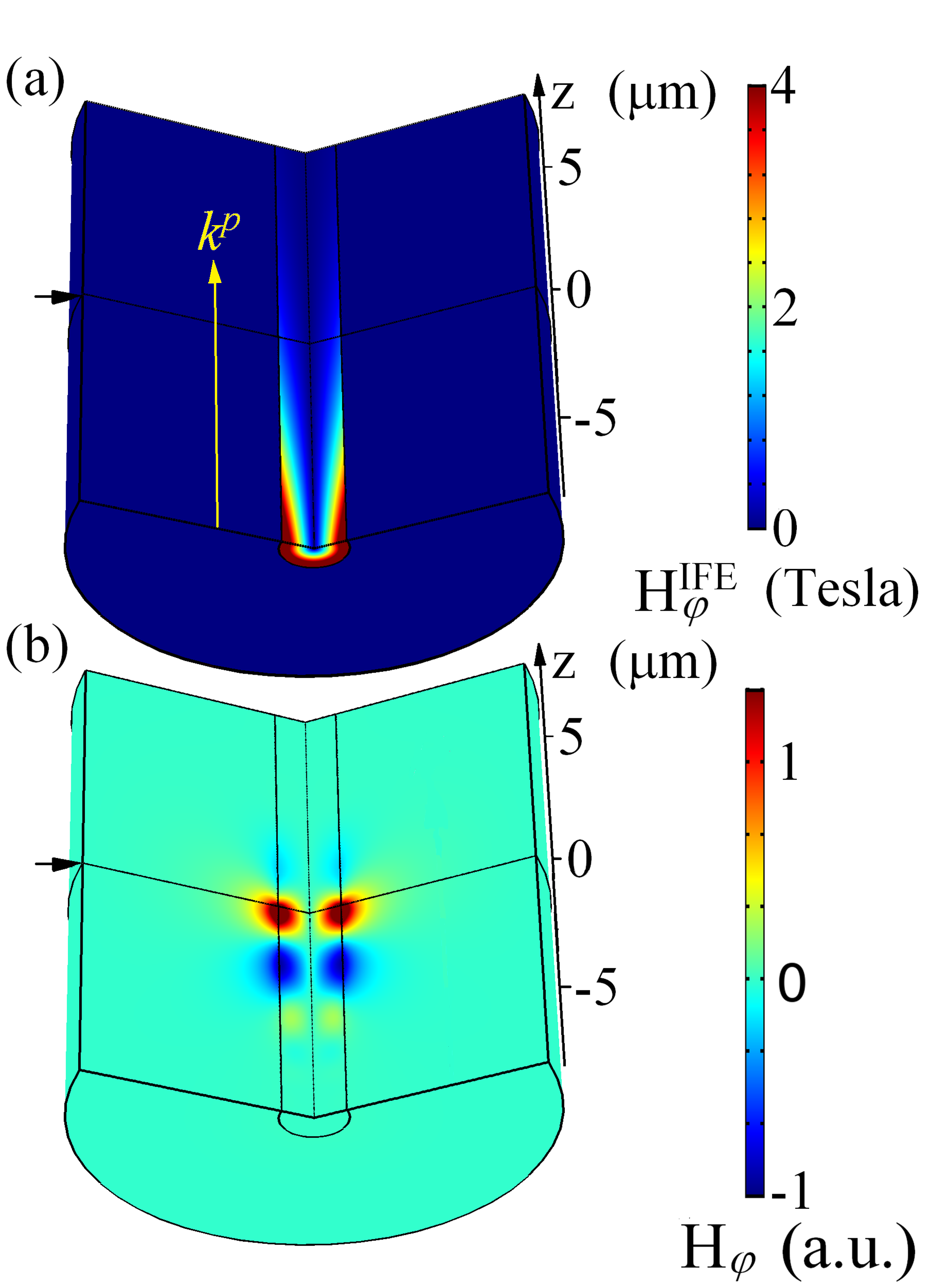}
\caption{Nonreciprocal propagation with a significant loss in the semiconductor wire. (a) Distribution of the static azimuthal magnetic field $H_\varphi^{\mathrm{IFE}}$ induced by the plasmonic pump power flow. (b) Distribution of the azimuthal magnetic field component of the plasmonic signal. A pump frequency of $\omega^\mathrm{p}=0.57\omega_\mathrm{pl}$, a signal frequency of $\omega^\mathrm{s}=0.76\omega_\mathrm{pl}$, an input pump power of $P$=300W and a significant loss of $\gamma_s=0.11\omega_\mathrm{pl}$ have been used.}
\label{fig:4}
\end{figure}

A plasmonic pump power flow is carried by the fundamental plasmonic mode of the semiconductor wire. The fundamental plasmonic mode is strongly confined at the surface of the semiconductor wire, as shown in Fig. 1(b).  The fundamental plasmonic mode has a rotating electric field vector which is manifested by $\pi$-difference in phase between the $\hat{\rho}$-axis electric field component $E_\rho^\mathrm{p}$  and the $\hat{z}$-axis electric field component $E_z^\mathrm{p}$ as shown in Fig. 1(c). The rotating electric field vector of the high-power pump acts as a static magnetic field along the azimuthal direction by the IFE which is described as
\begin{eqnarray}
{\vec{H}}^{\mathrm{IFE}} = i\alpha \left( {{\omega ^{\rm{p}}}} \right){\varepsilon _0}/{\mu _0}\left( {E_\rho ^{\rm{p}}E{{_z^{\rm{p}}}^ * } - E_z^{\rm{p}}E{{_\rho ^{\rm{p}}}^ * }} \right)\hat{\varphi},
\label{eq2}
\end{eqnarray}
where the superscript p represents the pump. Strength of the optically induced magnetic field decreases with increasing the wire radius as shown in Fig. 1(d) and increases with increasing the pump frequency as shown in Fig. 1(e), while keeping a pump power of $P$=1W. 

From Eqs. (1) and (2), we can get the dielectric permittivity tensor.
\begin{eqnarray}
\centering
\begin{array}{l}
\left( {\begin{array}{*{20}{c}}
{{D_\rho }\hat \rho }\\
{{D_\varphi }\hat \varphi }\\
{{D_z}\hat z}
\end{array}} \right) = \\
{\varepsilon _0}\left( {\begin{array}{*{20}{c}}
{{\varepsilon _1}}&0&{i\alpha H_\varphi ^{\mathrm{IFE}}}\\
0&{{\varepsilon _1}}&0\\
{ - i\alpha H_\varphi ^{\mathrm{IFE}}}&0&{{\varepsilon _1}}
\end{array}} \right)\left( {\begin{array}{*{20}{c}}
{{E_\rho }\hat \rho }\\
{{E_\varphi }\hat \varphi }\\
{{E_z}\hat z}
\end{array}} \right).
\end{array}
\label{eq3}
\end{eqnarray}
Here $\alpha H_\varphi^{\mathrm{IFE}}$ of the off-diagonal component of the dielectric permittivity tensor is proportional to the azimuthal magnetic field induced by the plasmonic pump power flow. The optically induced azimuthal magnetic field alters the material parameters and thus the plasmonic signal, which can be considered as a type of third-order nonlinear optical effect.
Now we consider plasmonic signal propagation under the optically induced azimuthal magnetic field. To obtain the third order nonlinear susceptibility we start from a general form of the Lorentz reciprocity theorem (the detail derivation can be seen in Appendix A)
\begin{eqnarray}
\begin{array}{l}
\frac{\partial }{{\partial z}}\int {\left[ {{{\vec E}_1}\left( {\vec r} \right) \times {{\vec H}_2}\left( {\vec r} \right) - {{\vec E}_2}\left( {\vec r} \right) \times {{\vec H}_1}\left( {\vec r} \right)} \right]\cdot\hat zd\sigma }  = \\
i\omega \int {\left( {{{\vec E}_1}\left( {\vec r} \right)\cdot{{\vec D}_2}\left( {\vec r} \right) - {{\vec E}_2}\left( {\vec r} \right)\cdot{{\vec D}_1}\left( {\vec r} \right)} \right)} d\sigma,
\end{array}
\label{eq4}
\end{eqnarray}
where $\left({\vec{E}}_1,\ {\vec{H}}_1\right)$ and $\left({\vec{E}}_2,\ {\vec{H}}_2\right)$ are two arbitrary guided modes. The integration according to $\sigma$ denotes the integration on the two-dimensional area $({\rho}, {\varphi})$ perpendicular to the $\hat z$-axis. The integration domain is the mode confined area.  Now we substitute for $({\vec{E}}_1,\ {\vec{H}}_1)$ and $({\vec{E}}_2,\ {\vec{H}}_2)$ counterpropagating unperturbed fundamental mode and  perturbed fundamental mode under the optically induced azimuthal magnetic field $H_\varphi^{\mathrm{IFE}}$, respectively. In the first order of perturbation by the optically induced azimuthal magnetic field $H_\varphi^{\mathrm{IFE}}$ we get
\begin{eqnarray}
\Delta {k} =  - \frac{{i{k_0}\alpha \left( {{\omega ^{\rm{s}}}} \right) \int {E_\rho ^\mathrm{s}\;E_z^\mathrm{s}H_\varphi ^{\mathrm{IFE}}ds} }}{{{\eta _0}\int {E_\rho ^\mathrm{s}H_\varphi ^\mathrm{s}ds} }}
\label{eq5}
\end{eqnarray}
If we substitute Eq. (2) to Eq. (5) we get the cross-nonlinear susceptibility
\begin{eqnarray}
\nonumber
\gamma  = \frac{{\Delta {k}}}{P} = \\
\frac{{2{k_0}{\alpha \left( {{\omega ^{\rm{s}}}} \right)\alpha \left( {{\omega ^{\rm{P}}}} \right)}\int {E_\rho ^\mathrm{s}\;E_z^\mathrm{s}(E_\rho ^\mathrm{p}E{{_z^\mathrm{p}}^ * } - E_z^\mathrm{p}E{{_\rho ^\mathrm{p}}^ * })ds} }}{{{\eta _0}^3\int {E_\rho ^\mathrm{s}H_\varphi ^\mathrm{s}ds} \left| {\int {{\mathrm{Re}}\left( {E_\rho ^\mathrm{p}H{{_\varphi ^\mathrm{p}}^ * }} \right)ds} } \right|}}.
\label{eq6}
\end{eqnarray}
Here $k_0=\omega^\mathrm{s}/c$ is the vacuum wavenumber of the signal, $\eta_0\approx377\Omega$ is the vacuum wave impedance and $P=1/2\left|\int{{\rm{Re}}{\left(E_\rho^\mathrm{p}{H_\varphi^\mathrm{p}}^\ast\right)}ds}\right|\ $ is the mode power of the pump. The detail derivations of Eqs. (5) and (6) can be seen in Appendix B.
Eqs. (5) and (6) give quantitatively the optically-induced wavenumber shift and the cross-nonlinear susceptibility. From Eqs. (5) and (6) it is predicted that the optically-induced wavenumber shift has opposite signs for forward and backward pump power flows. Consequently, it is predicted that for a forward pump power flow forward and backward signals present opposite signs of wavenumber shift manifesting optically-induced nonreciprocity.

Fig. 2 shows dispersion relation of the fundamental plasmonic mode in the semiconductor wire carrying a plasmonic pump power flow. In the absence of the plasmonic pump power flow the fundamental plasmonic mode has been obtained by numerically solving the dispersion relation in cylindrical coordinates  \cite{39Davoyan2014, 40Bliokh2018} (the blue curves). In the presence of plasmonic pump power flow the wavenumber shift has been calculated by using Eq. (6). (the red curves). The red curves show that the wavenumber shift has opposite signs for forward signal (positive values of $k^\mathrm{s}$) and backward signal (negative values of $k^\mathrm{s}$) manifesting the nonreciprocal property as predicted in Eq. (6) while the blue curves present the reciprocal property.
With increasing the pump power flow the nonreciprocity becomes stronger. The results by Eq. (6) (the red curves) deviate from the the full simulation results (the red circles) near the SPP resonances where the wavenumbers have very large values because here an effect of the optically induced azimuthal magnetic field $H_\varphi^{\mathrm{IFE}}$ cannot be treated as a small perturbation. For $\omega^\mathrm{p}=0.73\omega_{\mathrm{pl}}$ (e) the nonreciprocity is stronger than that for $\omega^{\mathrm{p}}=0.63\omega_{\mathrm{pl}}$ (b) in spite of the same pump power flow because the higher pump frequency provides a stronger enhancement of $H_\varphi^{\mathrm{IFE}}$ due to a tighter mode confinement of the plasmonic pump. The nonlinear wavenumber shift is sign-reversed by changing the direction of the pump power flow (f). 
The condition that the cyclotron frequency $\omega_c$ must be much smaller than the optical signal frequency $\omega$ gives an upper limit on the strength of the pump field. For a pump mode power of 1W the maximum value of $\omega_c=0.0097\omega_{\rm{pl}}$ is much smaller than the optical signal frequency $\omega$, thus the above condition is well satisfied. On the other hand, the signal frequency should be near the plasmon resonance and the pump frequency should be away from the resonance because guiding plasmons near the resonance are very sensitive on an external magnetic field and those away from the resonance are not sensitive on the external magnetic field keeping a moderately low loss to influence the signal effectively. For the considered plasmonic structure, a pump mode power of 1W corresponds an intensity of about $10^7$W/cm$^2$ and a fluence of 0.01mJ/cm$^2$ for a picosecond pulse which is a typical value in experiments and much smaller than the damage threshold of solid-state materials \cite{10Kirilyuk2010}.  

In Fig. 3 and 4 we demonstrated the optically induced nonreciprocity with numerical simulations. Fig. 3 (a) shows the static azimuthal magnetic field induced by the plasmonic pump power flow. Fig. 3(b) shows the all-optical diode function based on the mode cutoff for the forward signal near the SPP resonances which was predicted in Fig. 2(c). A significant loss in the wire has been introduced in Fig. 4 because the loss reduces the performance of many of plasmonic devices significantly.

We discuss that our findings are promising for all-optical on-chip nonreciprocal devices based on simple waveguide structures. There have been intensive studies on optical nonreciprocity because of its practical importance. The magnetic field induced nonreciprocity is based on the asymmetric property of non-diagonal permittivity of materials under an external magnetic field, thus it has no further requirement on the structure. However, this conventional approach requires a bulky configuration for excitation of the external magnetic field rendering it infeasible for systems integration. Although the optically induced nonreciprocity is more appropriate for integrated photonic circuits and all-optical processing, this approach generally requires more complicated structures for achieving asymmetric field distributions. The proposed conception in this paper combines both of the advances of the magnetic field induced nonreciprocity and the optically induced nonreciprocity. Here, we note that chip compatible utilization of the IFE is not trivial. Most of already proposed configurations for the IFE require the free-space coupling scheme, thus are not compatible with on-chip integration. While the plasmonic waveguide structures are promising for on-chip integration with conventional dielectric waveguides, a feature necessary for future photonic nanocircuits, we utilized the recently studied scheme of the IFE in waveguide structures \cite{21Im2017}.

We found that semiconductor wires, which provide a very simple and efficient platform, are spatially asymmetry-free, however are naturally biased by a plasmonic pump power flow resulting in the nonreciprocal properties of the plasmonic propagation. The rotating electric field vector of the plasmonic pump in the semiconductor wires act as a static magnetic field along the azimuthal direction. The static azimuthal magnetic field induces a nonlinear wavenumber shift of the fundamental plasmonic mode which has opposite signs for forward and backward signals. With numerical simulations we demonstrated the all-optical diode function exploiting the mode cutoff for the forward signal near the SPP resonances. We note that also in states far away from the SPP resonances small differences of the nonlinear wavenumber shift for forward and backward signals can be used to provide the all-optical diode function with the help of a Mach-Zehnder interferometer. Our findings may open an avenue towards highly integrated all-optical diodes.

\appendix

\begin{widetext}

\section{Derivations of Eq. (4)}

Consider two arbitrary modes with the fields ${\vec{E}}_1\left(\vec{r}\right),\ {\vec{H}}_1\left(\vec{r}\right),{\vec{E}}_2\left(\vec{r}\right)$ and ${\vec{H}}_2\left(\vec{r}\right)$.
The time-harmonic sourceless Maxwell equations for the first mode are
\begin{eqnarray}
\mathrm{\nabla}\times{\vec{E}}_1=i\omega{\vec{B}}_1
\label{eqs1}
\end{eqnarray}  
\begin{eqnarray}
\nabla\times{\vec{H}}_1=-i\omega{\vec{D}}_1.
\label{eqs2}
\end{eqnarray} 
Dot multiplying Eq. (A1) with $H_2$ and Eq. (A2) with $E_2$ and then summing gives 
\begin{eqnarray}
{\vec{H}}_2\mathrm{\nabla}\times{\vec{E}}_1+{\vec{E}}_2\mathrm{\nabla}\times{\vec{H}}_1=-i\omega\left({\vec{E}}_2{\vec{D}}_1-{\vec{H}}_2{\vec{B}}_1\right).
\label{eqs3}
\end{eqnarray} 
Applying the same process with interchanged primes yields
\begin{eqnarray}
{\vec{H}}_1\mathrm{\nabla}\times{\vec{E}}_2+{\vec{E}}_1\mathrm{\nabla}\times{\vec{H}}_2=-i\omega\left({\vec{E}}_1{\vec{D}}_2-{\vec{H}}_1{\vec{B}}_2\right).
\label{eqs4}
\end{eqnarray} 
Subtracting these two equations we obtain
\begin{eqnarray}
\mathrm{\nabla}\cdot\left({\vec{E}}_1\times{\vec{H}}_2-{\vec{E}}_2\times{\vec{H}}_1\right)=-i\omega\left({\vec{E}}_2{\vec{D}}_1-{\vec{E}}_1{\vec{D}}_2-{\vec{H}}_2{\vec{B}}_1+{\vec{H}}_1{\vec{B}}_2\right).
\label{eqs5}
\end{eqnarray}
If we remind that the time-harmonic fields have been assumed, at the optical frequency $\omega$  
\begin{eqnarray}
\nonumber
{\vec{H}}_2(\omega){\vec{B}}_1(\omega)={\vec{H}}_2(\omega){\hat{\mu}}_1(\omega){\vec{H}}_1(\omega),\\
\nonumber
{\vec{H}}_1(\omega){\vec{B}}_2(\omega)={\vec{H}}_1(\omega){\hat{\mu}}_2(\omega){\vec{H}}_2(\omega).
\label{eqs51}
\end{eqnarray}
At the optical frequency $\omega$, ${\hat{\mu}}_1(\omega)={\hat{\mu}}_2(\omega)=\mu_0{\hat{I}},$ where ${\hat{I}}$ is the unit tensor.
Here, we note that an effect of the static magnetic field ${\vec{H}}^{\mathrm{IFE}}$ by the IFE is not considered for ${\hat{\mu}}(\omega)$ and ${\vec{H}}(\omega)$ at the optical frequency. An effect of the static magnetic field ${\vec{H}}^{\mathrm{IFE}}$ has been considered for the dielectric permittivity tensor ${\hat{\varepsilon}}(\omega)$ because the off-diagonal component of the dielectric permittivity tensor is proportional to the static magnetic field ${\vec{H}}^{\mathrm{IFE}}$ as shown in Eq. (3).
Therefore, ${\vec{H}}_2(\omega){\vec{B}}_1(\omega)={\vec{H}}_1(\omega){\vec{B}}_2(\omega)$ and
\begin{eqnarray}
\mathrm{\nabla}\cdot\left({\vec{E}}_1\times{\vec{H}}_2-{\vec{E}}_2\times{\vec{H}}_1\right)=-i\omega\left({\vec{E}}_2{\vec{D}}_1-{\vec{E}}_1{\vec{D}}_2\right).
\label{eqs6}
\end{eqnarray}
Integration of the both term of Eq. (A6) in the volume between two planes at $z$ and $z+\mathrm{\Delta z}$ gives
\begin{eqnarray}
\int{\mathrm{\nabla}\cdot\left({\vec{E}}_1\times{\vec{H}}_2-{\vec{E}}_2\times{\vec{H}}_1\right)dV}=-i\omega\int\left({\vec{E}}_2{\vec{D}}_1-{\vec{E}}_1{\vec{D}}_2\right)dV.
\label{eqs7}
\end{eqnarray}
Using the divergence theorem we obtain
\begin{eqnarray}
\oint{\left({\vec{E}}_1\times{\vec{H}}_2-{\vec{E}}_2\times{\vec{H}}_1\right)d\vec{\sigma}}=-i\omega\int\left({\vec{E}}_2{\vec{D}}_1-{\vec{E}}_1{\vec{D}}_2\right)dV.
\label{eqs8}
\end{eqnarray}
If $\mathrm{\Delta z}$ converges to 0, Eq. (A8) can be rewritten as
\begin{eqnarray}
\nonumber
\int\limits_{\sigma \left( {z + \Delta z} \right)} {\left( {{{\vec E}_1} \times {{\vec H}_2} - {{\vec E}_2} \times {{\vec H}_1}} \right)\cdot\vec zd\sigma }  - \int\limits_{\sigma \left( z \right)} {\left( {{{\vec E}_1} \times {{\vec H}_2} - {{\vec E}_2} \times {{\vec H}_1}} \right)\cdot\vec zd\sigma }  =  - i\omega \Delta z\int {\left( {{{\vec E}_2}{{\vec D}_1} - {{\vec E}_1}{{\vec D}_2}} \right)} d\sigma . 
\label{eqs9}
\end{eqnarray}
Rewriting the above equation gives
\begin{eqnarray}
\frac{\partial}{\partial z}\int{\left({\vec{E}}_1\times{\vec{H}}_2-{\vec{E}}_2\times{\vec{H}}_1\right)\cdot\vec{z}d\sigma}=i\omega\int\left({\vec{E}}_1{\vec{D}}_2-{\vec{E}}_2{\vec{D}}_1\right)d\sigma.	
\label{eqs10}
\end{eqnarray}
\\

\section{Derivations of Eqs. (5) and (6)}

Now we substitute for $\left({\vec{E}}_1,{\vec{H}}_1\right)$ and $\left({\vec{E}}_2,{\vec{H}}_2\right)$ the unperturbed backward-propagating field $\left({\vec{E}}^-,{\vec{H}}^-\right)$ and the perturbed forward-propagating field $\left({\vec{E}}^+,{\vec{H}}^+\right)$ under an external transverse magnetic field, respectively.
The external magnetic field induces a magnetic field ${\vec{H}}^{{\rm{IFE}}}$ in the $\varphi$ -direction and leads to a perturbation for the mode distribution
\begin{eqnarray}
\begin{array}{l}
{{\vec E}^ - } = {{\vec E}^ - }(\rho ,\varphi )\exp ( - ikz),\\
{{\vec H}^ - } = {{\vec H}^ - }(\rho ,\varphi )\exp ( - ikz),\\
{{\vec E}^ + } = \left( {{{\vec E}^ + }\left( {\rho ,\varphi } \right) + \Delta {{\vec E}^ + }\left( {\rho ,\varphi } \right)} \right){\rm{exp}}\left( {i\left( {k + \Delta k} \right)z} \right),\\
{{\vec H}^ + } = \left( {{{\vec H}^ + }\left( {\rho ,\varphi } \right) + \Delta {{\vec H}^ + }\left( {\rho ,\varphi } \right)} \right){\rm{exp}}\left( {i\left( {k + \Delta k} \right)z} \right).
\end{array}
\label{eqs2}
\end{eqnarray}
From a physical insight on backward- and forward-propagating fields, the components of electric field and magnetic field satisfy
$E_z^-\left(\rho,\varphi\right)=-E_z^+\left(\rho,\varphi\right)$, $E_\varphi^-\left(\rho,\varphi\right)=E_\varphi^+\left(\rho,\varphi\right)$, $E_\rho^-\left(\rho,\varphi\right)=E_\rho^+\left(\rho,\varphi\right)$, $H_z^-\left(\rho,\varphi\right)=H_z^+\left(\rho,\varphi\right)$, $H_\varphi^-\left(\rho,\varphi\right)=-H_\varphi^+\left(\rho,\varphi\right)$ and $H_\rho^-\left(\rho,\varphi\right)=-H_\rho^+\left(\rho,\varphi\right)$.
From the above relations we derive
\begin{eqnarray}
{\vec{E}}^-\left(\rho,\varphi\right)\times{\vec{H}}^+\left(\rho,\varphi\right)-{\vec{E}}^+\left(\rho,\varphi\right)\times{\vec{H}}^-\left(\rho,\varphi\right)=2\left({\vec{E}}^+\left(\rho,\varphi\right)\times{\vec{H}}^+\left(\rho,\varphi\right)\right).
\label{eqs11}
\end{eqnarray}

Using Eq. (B2) we can express the left side of Eq. (A9) in terms of ${{E}}_\rho^+\left(\rho,\varphi\right)$ and ${{H}}_\varphi^+\left(\rho,\varphi\right)$, 
\begin{eqnarray}
\nonumber
{\frac{\partial }{{\partial z}}\int {\left( {{{\vec E}^ - } \times {{\vec H}^ + } - {{\vec E}^ + } \times {{\vec H}^ - }} \right)\cdot\vec zd\sigma } }\\
\nonumber
{ = \frac{\partial }{{\partial z}}\int {\exp \left( {i\Delta k{\rm{z}}} \right)\left( {{{\vec E}^ - }\left( {\rho ,\varphi } \right) \times \left( {{{\vec H}^ + }\left( {\rho ,\varphi } \right) + \Delta {{\vec H}^ + }\left( {\rho ,\varphi } \right)} \right)} \right)\cdot\vec zd\sigma } }\\
\nonumber
{ - \frac{\partial }{{\partial z}}\int {\exp \left( {i\Delta k{\rm{z}}} \right)\left( {\left( {{{\vec E}^ + }\left( {\rho ,\varphi } \right) + \Delta {{\vec E}^ + }\left( {\rho ,\varphi } \right)} \right) \times {{\vec H}^ - }\left( {\rho ,\varphi } \right)} \right)\cdot\vec zd\sigma } }\\
\nonumber
{ \approx \frac{\partial }{{\partial z}}\int {{\rm{exp}}\left( {i\Delta k{\rm{z}}} \right)\left( {{{\vec E}^ - }\left( {\rho ,\varphi } \right) \times {{\vec H}^ + }\left( {\rho ,\varphi } \right) - {{\vec E}^ + }\left( {\rho ,\varphi } \right) \times {{\vec H}^ - }\left( {\rho ,\varphi } \right)} \right)\cdot\vec zd\sigma } }\\
\nonumber
{ = 2i\Delta k{\rm{exp}}\left( {i\Delta k{\rm{z}}} \right)\int {\left( {{{\vec E}^ + }\left( {\rho ,\varphi } \right) \times {{\vec H}^ + }\left( {\rho ,\varphi } \right)} \right)\cdot\vec zd\sigma } }\\
\nonumber
{ = 2i\Delta k{\rm{exp}}\left( {i\Delta k{\rm{z}}} \right)\int { E_\rho ^ + \left( {\rho ,\varphi } \right) H_\varphi ^ + \left( {\rho ,\varphi } \right)d\sigma } }.
\label{eqs12}
\end{eqnarray}

Now in the right side of Eq. (A9) the electric displacements are expressed as
\begin{eqnarray}
\begin{array}{l}
{{\vec D}_1} = {\varepsilon _0}\varepsilon {{\vec E}^ - },\\
{{\vec D}_2} = {\varepsilon _0}\hat \varepsilon {{\vec E}^ + } = {\varepsilon _0}\left( {\varepsilon {{\vec E}^ + } + i\alpha {{\vec E}^ + } \times {{\vec H}^{\rm{IFE}}}} \right).
\end{array}
\label{eqs13}
\end{eqnarray}
Using Eq. (B3) we can express the right side of Eq. (A9) in terms of ${{E}}_\rho^+\left(\rho,\varphi\right)$ and ${{H}}_\varphi^+\left(\rho,\varphi\right)$,
\begin{eqnarray}
\nonumber
i\omega \int {\left( {{{\vec E}^ - }{{\vec D}^ + } - {{\vec E}^ + }{{\vec D}^ - }} \right)} d\sigma 
 = i\omega \int {\left( {{{\vec E}^ - }{\varepsilon _0}\left( {{{\vec D}^ + } + i\alpha {{\vec E}^ + } \times {{\vec H}^{{\rm{IFE}}}}} \right) - {{\vec E}^ + }{\varepsilon _0}{{\vec D}^ - }} \right)} d\sigma \\
\nonumber
 =  - \frac{{{k_0}\alpha }}{{{\eta _0}}}\int {{{\vec E}^ - }\left( {{{\vec E}^ + } \times {{\vec H}^{{\rm{IFE}}}}} \right)d\sigma } 
 =  - \frac{{{k_0}\alpha }}{{{\eta _0}}}{\rm{exp}}\left( {i\Delta k{\rm{z}}} \right)\int {\left( { E_z^ - \left( {\rho ,\varphi } \right) E_\rho ^ + \left( {\rho ,\varphi } \right){{ H}^{{\rm{IFE}}}} -  E_\rho ^ - \left( {\rho ,\varphi } \right) E_z^ + \left( {\rho ,\varphi } \right){{ H}^{{\rm{IFE}}}}} \right)} d\sigma \\
\nonumber
 = \frac{{2{k_0}\alpha }}{{{\eta _0}}}{\rm{exp}}\left( {i\Delta k{\rm{z}}} \right)\int {{{ H}^{{\rm{IFE}}}}E_\rho ^ + \left( {\rho ,\varphi } \right)E_z^ + \left( {\rho ,\varphi } \right)d\sigma } .
\label{eqs2}
\end{eqnarray}

From the both sides of Eq. (A9) we obtain
\begin{eqnarray}
2i\Delta k{\rm{exp}}\left( {i\Delta k{\rm{z}}} \right)\int { E_\rho ^ + \left( {\rho ,\varphi } \right) H_\varphi ^ + \left( {\rho ,\varphi } \right)d\sigma }  = \frac{{2{k_0}\alpha }}{{{\eta _0}}}{\rm{exp}}\left( {i\Delta k{\rm{z}}} \right)\int {{{ H}^{{\rm{IFE}}}}E_\rho ^ + \left( {\rho ,\varphi } \right)E_z^ + \left( {\rho ,\varphi } \right)d\sigma } .
\label{eqs2}
\end{eqnarray}
If we substitute ${{E}}_\rho^{\rm{s}}, {{E}}_z^{\rm{s}}$ and ${{H}}_\varphi^{\rm{s}}$ for ${{E}}_\rho^+\left(\rho,\varphi\right), {{E}}_z^+\left(\rho,\varphi\right)$ and ${{H}}_\varphi^+\left(\rho,\varphi\right)$
we obtain	
\begin{eqnarray}
\Delta k =  - \frac{{i{k_0}\alpha \int {E_\rho ^{\rm{s}}E_z^{\rm{s}}{H^{{\rm{IFE}}}}d\sigma } }}{{{\eta _0}\int {E_\rho ^{\rm{s}}H_\varphi ^{\rm{s}}d\sigma } }}.
\label{eqs2}
\end{eqnarray}
The mode power is expressed as $P = 1/2\left| {\int {\rm{Re}}\left( {E_\rho ^{\rm{p}}H{{_\varphi ^{\rm{p}}}^ * }} \right)d\sigma }  \right|$ and we obtain
\begin{eqnarray}
\gamma  = \frac{{\Delta k}}{P} = \frac{{ - i2{k_0}\alpha \left( {{\omega ^{\rm{s}}}} \right)\int {E_\rho ^{\rm{s}}E_z^{\rm{s}}{H^{{\rm{IFE}}}}d\sigma } }}{{{\eta _0}\int {E_\rho ^{\rm{s}}H_\varphi ^{\rm{s}}d\sigma } \left| {\int {\rm{Re}}\left( {E_\rho ^{\rm{p}}H{{_\varphi ^{\rm{p}}}^ * }} \right)d\sigma } \right|}} = \frac{{2{k_0}\alpha \left( {{\omega ^{\rm{s}}}} \right)\alpha \left( {{\omega ^{\rm{p}}}} \right)\int {E_\rho ^{\rm{s}}E_z^{\rm{s}}\left( {E_\rho ^{\rm{p}}E{{_z^{\rm{p}}}^ * } - E_z^{\rm{p}}E{{_\rho ^{\rm{p}}}^ * }} \right)d\sigma } }}{{{\eta _0}^3\int {E_\rho ^{\rm{s}}H_\varphi ^{\rm{s}}d\sigma } \left| {\int {\rm{Re}}\left( {E_\rho ^{\rm{p}}H{{_\varphi ^{\rm{p}}}^ * }} \right)d\sigma } \right|}},
\label{eqs2}\
\end{eqnarray}
where ${H^{{\rm{IFE}}}} = i\alpha \left( {{\omega ^{\rm{p}}}} \right){\varepsilon _0}/{\mu _0}\left( {E_\rho ^{\rm{p}}E{{_z^{\rm{p}}}^ * } - E_z^{\rm{p}}E{{_\rho ^{\rm{p}}}^ * }} \right)$.

\end{widetext}

\bibliography{nonreciprocity}
\end{document}